\documentstyle[prl,aps,twoside,floats,epsf]{revtex}

\begin{document}
\draft

\twocolumn[\hsize\textwidth\columnwidth\hsize\csname
@twocolumnfalse\endcsname
\draft 
\title{
Angular dependence of 
novel magnetic quantum oscillations in 
a quasi-two-dimensional multiband  Fermi liquid
with impurities 
}
\author{ A.M. Bratkovsky$^1$ and  A.S. Alexandrov$^2$  }
\address{$^1$ Hewlett-Packard Laboratories, 1501 Page Mill Road, 
Palo Alto, California 94304}
\address{ 
$^2$ Department of Physics, Loughborough University, LE11 3TU, United
Kingdom 
}
\date{April 24, 2001 }
\maketitle
\begin{abstract}
The semiclassical Lifshitz-Kosevich-type description 
is given for the angular dependence of quantum oscillations with 
combination frequencies in 
a multiband quasi-two-dimensional  Fermi liquid with a constant 
number of electrons.  The analytical  expressions are found 
for the Dingle, thermal, spin,  and amplitude (Yamaji) reduction
factors of the novel combination harmonics,  
where the latter two strongly oscillate  with the direction  of the field.
At the ``magic" angles those factors reduce to the purely two-dimensional
expressions given earlier. The combination harmonics are suppressed 
in the presence of the non-quantized (``background") states, 
and they decay 
exponentially faster with temperature and/or disorder compared 
to the standard harmonics, 
providing an additional tool for electronic structure determination.
The theory is applied to Sr$_2$RuO$_4$.

\end{abstract}
\vskip 2pc ]
\narrowtext

The low-dimensional organic conductors exhibit very unusual electronic
properties, like magnetic field induced superconductivity, and are the focus
of solid state research\cite{uji2001}. Magnetic oscillations of
susceptibility and resistivity are the main source of detailed information
about their electronic structure, yet there is no complete theory of these
phenomena for quasi-two-dimensional (quasi-2D) metals. The analytical
semiclassical (Lifshitz-Kosevich type\cite{david}) theory of the de Haas-van
Alphen combination harmonics in the multiband two-dimensional (2D)\ metals has
been suggested recently \cite{ABprb01}. Even earlier it was shown \cite
{alebra} for a system like Sr$_{2}$RuO$_{4}$ \cite{andy96}, that the magnetic
quantum oscillations in a multiband 2D metal with a fixed electron density
[canonical ensemble (${\rm CE}$)] are qualitatively different from those in
an open system where the chemical potential is fixed [grand canonical
ensemble (${\rm GCE}$)]. The chemical potential oscillates with the field in
closed systems and this provides a mechanism for different bands to
communicate with each other in ${\rm CE}$ producing a dHvA signal with the
``sum'' \cite{alebra} and the ``difference'' combination frequencies\cite
{alebra1,nakano,nakano99} in addition to the ordinary dHvA frequencies$.$
Those novel oscillations have been observed in quantum well structures \cite
{shepherd99} and Sr$_{2}$RuO$_{4}$\cite{yoshida98,ohmichi99}. Their
amplitudes are comparable with the standard components, and they are robust
with respect to any {\em background} (non-quantized) density of states at
low temperatures \cite{alebra1}, but fall off exponentially faster with the
temperature \cite{ABprb01}.

Many layered systems, including Sr$_{2}$RuO$_{4},$ are actually {\em quasi}%
-2D. Weak coupling between conducting layers, which introduces a dispersion
of the Fermi surface in the direction perpendicular to the layers, leads to
a strong dependence of magnetic oscillations on the angle $\theta $\ between
the normal to the conducting planes and the magnetic field $\vec{B}$\cite
{shchegolev}. This is related to the fact that the energy spectrum in the
field becomes almost 2D (the width of the Landau minibands almost vanishes)
at some ``magic'' (Yamaji) angles\cite{yamaji}, where the amplitude of the
oscillations is strongly enhanced. The theory \cite{ABprb01} applies only to
the clean systems at the magic angles. The full analytical semiclassical theory of
the quasi-2D multiband canonical metals at finite temperature with full account
for the angular dependence of the Landau miniband widths and the spin factors,
the collision broadening of the minibands (Dingle factor), and the
background density of states is given in the present paper and applied to
a generic case of Sr$_{2}$RuO$_{4}.$

First, we derive a convenient expression for a multiband two-dimensional
thermodynamic potential in magnetic field $B$, mainly in the units $\hbar
=\left| e\right| =c=1,$ 
\begin{equation}
\Omega =-T\int d\epsilon {\cal N}(\epsilon ,B)\ln \left( 1+\exp {\frac{\mu
-\epsilon }{{T}}}\right)  \label{eq:Omega}
\end{equation}
Consider the multiband quasi-2D\ system with a general dispersion law 
\begin{equation}
\epsilon =\Delta _{\alpha }+\frac{k_{x}^{2}+k_{y}^{2}}{2m_{\alpha }}%
-2t_{\alpha }\cos jk_{z}d,  \label{eq:Ecos}
\end{equation}
where $k_{x},k_{y}$ are the momenta in the conducting plane, $t_{\alpha }$
is the hopping between the (conducting) layers, $d$ the distance between the
layers, $j$ the integer number. We are mainly interested in the situation
where the kinetic energy of electrons is much larger than the hopping
between layers, $\mu -\Delta _{\alpha }\gg 2t_{\alpha }.$ The density of
states (DOS) with an account for collision broadening of the Landau levels
(Dingle factor)\cite{dingle,bychkov} can be written as 
\begin{eqnarray}
{\cal N}(\epsilon ,B) &=&\sum_{\alpha ,bg}^{{\rm bands,spin}}{\cal N}%
_{\alpha }(\epsilon ,B),  \label{eq:sdos} \\
{\cal N}_{\alpha }(\epsilon ,B) &=&-\frac{1}{\pi }%
\mathop{\rm Im}%
\sum_{n=0}^{\infty }\frac{\Delta S}{\left( 2\pi \right) ^{3}}\int_{-\pi
/d}^{\pi /d}\frac{dk_{z}}{\epsilon -\epsilon _{\alpha ,nk_{z}+i\Gamma
_{\alpha }}},  \label{eq:Nados}
\end{eqnarray}
where $\Delta S=2\pi |e|B/\hbar c$ is the cross-sectional area in k-space
between the two successive Landau orbits, 
\begin{equation}
\epsilon _{\alpha ,nk_{z}}=\Delta _{\alpha 0}+\omega _{\alpha
}(n+1/2)+D_{\alpha }\cos jk_{z}d+g_{\alpha }\sigma \mu _{B}B,
\label{eq:EnkzY}
\end{equation}
the energy dispersion for a present geometry of the Landau orbits,
where $D_{\alpha }=2t_{\alpha }J_{0}\left( jk_{f\alpha }d\tan \theta
\right)$  with $J_{0}(x)$ the zeroth-order Bessel function, 
$\Gamma_\alpha =\pi /2\tau _{\alpha },$ $\tau _{\alpha }$ the scattering
mean free time in zero field\cite{bychkov} in the band $\alpha ,$ and the
background ($bg)$ is included in the DOS (\ref{eq:sdos}) too\cite{alebra1}. After
integration we obtain an important exact result 
\begin{eqnarray}
{\cal N}_{\alpha }(\epsilon ,B) &=&\sum_{n=0}^{\infty }\rho _{\alpha }\omega
_{\alpha }\nu (\epsilon -\epsilon _{\alpha ,n}),  \label{eq:dosq} \\
\nu (\epsilon -\epsilon _{\alpha ,n}) &=&%
\mathop{\rm Im}%
\frac{i}{\pi \sqrt{D_{\alpha }^{2}-(\epsilon -\epsilon _{\alpha ,n}+i\Gamma
_{\alpha })^{2}}},
\end{eqnarray}
where $k_{f\alpha }^{2}/2m_{\alpha }=\mu -\Delta _{\alpha },$ 
\begin{equation}
\epsilon _{\alpha ,n}=\Delta _{\alpha 0}+\omega _{\alpha }(n+1/2)+g_{\alpha
}\sigma \mu _{B}B,
\end{equation}
$\omega _{\alpha }=\left| B\cos \theta /m_{\alpha }c\right| $ the cyclotron
frequency with the cyclotron mass $m_{\alpha }$, $\Delta _{\alpha 0}$ the
band edge in zero magnetic field, $\mu $ the chemical potential, $g_{\alpha
} $ the electron $g$-factor, $\sigma =\pm 1/2$, $\mu _{B}$ the Bohr
magneton. The band index $\alpha \equiv b\sigma $ includes the band index $b$
and the spin index $\sigma $. There $\rho _{\alpha }$ is the zero-field density
of states in the band $\alpha $. For the energies of interest, $\left| \epsilon
-\Delta _{\alpha }\right| >2t_{\alpha },$ $\rho _{\alpha }=m_{\alpha }/2\pi
\hbar ^{2}d$ per energy and unit volume, otherwise $\rho _{\alpha }=\left(
m_{\alpha }/2\pi ^{2}\hbar ^{2}d\right) \arccos [(\epsilon -\Delta
_{\alpha })/2t_{\alpha }].$ Note that in a clean system at the magic angles,
where $D_{\alpha }=\Gamma _{\alpha }=0,$ the expression for $\nu$
reduces to $\nu (\epsilon
-\epsilon _{\alpha ,n})=\delta \left( \epsilon -\epsilon _{\alpha ,n}\right)
,$ meaning that the spectrum becomes 2D, the Landau minibands reduce
to the Landau
levels, and the previous expressions \cite{ABprb01} fully apply. The
background density of states, $\rho _{bg},$ corresponding to possible
non-quantized (or largely broadened) bands, can be included in (\ref{eq:dosq}%
) as the formal limit $\omega _{bg}\rightarrow 0.$ Obviously, the
non-quantized background will contribute to the non-oscillating characteristics
of the system. In a clean limit the density of states takes a standard form
with the one-dimensional square-root singularities ${\cal N}\sim 1/\sqrt{%
D_{\alpha }^{2}-(\epsilon -\epsilon _{\alpha ,n})^{2}}$\ \cite{mik}.

By applying the Poisson formula \cite{david} to the sum over $n$ in 
the thermodynamic potential
\begin{eqnarray}
\Omega &=&-T\sum_{n=0}^{\infty }\int d\epsilon \sum_{\alpha ,bg}\rho
_{\alpha }\omega _{\alpha }\nu (\epsilon -\epsilon _{\alpha ,n})  \nonumber
\\
&&\times \ln \left( 1+\exp {\frac{\mu -\epsilon }{{T}}}\right)
\end{eqnarray}
with $\mu _{\alpha }=\mu -\Delta _{\alpha }$ and $\Delta _{\alpha }=\Delta
_{\alpha 0}+g_{\alpha }\sigma \mu _{B}B$, it can be written as 
\begin{equation}
\Omega =\Omega _{0}+\tilde{\Omega},
\end{equation}
where, after substituting $x=(\epsilon -\epsilon _{\alpha ,n})/D_{\alpha },$ 
\begin{eqnarray}
\Omega _{0} &=&-T\int_{0}^{\infty }d\epsilon \sum_{\alpha ,bg}\rho _{\alpha
}\int_{-\infty }^{\infty }\frac{dx}{\pi }%
\mathop{\rm Im}%
\frac{i}{\sqrt{1-\left( x+i\Gamma _{\alpha }^{\prime }\right) ^{2}}} 
\nonumber \\
&&\times \ln \left( 1+\exp {\frac{\mu _{\alpha }-\epsilon -D_{\alpha }x}{{T}}%
}\right)  \label{eq:O0}
\end{eqnarray}
is the ``classical'' part of the thermodynamic potential, with $\Gamma
_{\alpha }^{\prime }\equiv \Gamma _{\alpha }/D_{\alpha }$. In the GCE $\Omega
_{0}$ does not oscillate as a function of $1/B$, and contains the
contribution due to spin susceptibility (Pauli paramagnetism). At low
temperatures one finds 
\begin{equation}
\Omega _{0}=-\sum_{\alpha ,bg}\frac{1}{2}\rho _{\alpha }\left( \mu _{\alpha
}^{2}+\frac{1}{2}D_{\alpha }^{2}\right) .  \label{eq:Om0}
\end{equation}

The oscillating part of the thermodynamic potential includes only the Landau
quantized bands, 
\begin{eqnarray}
\tilde{\Omega} &=&-2T\sum_{\alpha }\rho _{\alpha }\sum_{r=1}^{\infty
}\int_{0}^{\infty }d\epsilon \int_{-\infty }^{\infty }\frac{dx}{\pi }%
\mathop{\rm Im}%
\frac{i}{\sqrt{1-\left( x+i\Gamma _{\alpha }^{\prime }\right) ^{2}}} 
\nonumber \\
&&\times \ln \left( 1+\exp {\frac{\mu _{\alpha }-\epsilon -D_{\alpha }x}{{T}}%
}\right) \cos 2\pi r\left( \frac{\epsilon }{\omega _{\alpha }}-\frac{1}{2}%
\right) .  \label{eq:Otint}
\end{eqnarray}
This expression, after integrating over $\epsilon $ and $x,$ reduces to 
\begin{eqnarray}
\tilde{\Omega} &=&{\frac{1}{{24}}}\sum_{\alpha }\rho _{\alpha }\omega
_{\alpha }^{2}+2\sum_{\alpha }\sum_{r=1}^{\infty }A_{\alpha }^{\prime r}\cos
2\pi r\left( {\frac{F_{\alpha }}{{B}}-}\frac{1}{2}-\psi _{\alpha }\right) 
\nonumber \\
&=&{\frac{1}{{24}}}\sum_{\alpha }\rho _{\alpha }\omega _{\alpha
}^{2}+4\sum_{b}^{{\rm bands}}\sum_{r=1}^{\infty }A_{b}^{r}\cos 2\pi r\left( {%
\frac{F_{b}}{{B}}-}\frac{1}{2}\right) ,  \label{eq:Ot}
\end{eqnarray}
where $F_{b}=\left( \mu -\Delta _{b0}\right) |m_{b}|/\left( 2\mu _{B}m\cos
\theta \right) \equiv \hbar cS_{fb}/2\pi e,$ $S_{fb}$ is the mean Fermi
surface zero-field cross-section (in standard units), and the spin-related
phase $\psi _{\alpha }=\sigma \gamma _{b},$ $\gamma _{b}=g_{b}\left|
m_{b}\right| /(2m\cos \theta )$. If the scattering time were not
dependent on the spin projection, the phase would be the only quantity
explicitly depending on spin in Eq.~(\ref{eq:Ot}), hence one can perform
a summation over $\sigma $ in the second term to reveal the standard spin
reduction factor $\cos \pi r\gamma _{b}$\cite{david}.

The amplitudes of the Fourier harmonics in Eq. (\ref{eq:Ot}) are
explicitly given by
\begin{equation}
A_{b}^{r}=\frac{\rho _{b}\omega _{b}^{2}}{4\pi ^{2}r^{2}}R_{T}\left( \frac{{%
2\pi ^{2}rT}}{{\omega _{b}}}\right) R_{Y}\left( \frac{2\pi rD_{b}}{\omega
_{b}}\right) R_{D}R_{s},  \label{eq:Ab}
\end{equation}
where $R_{T}\left( z\right) =z/\sinh z$ \ is the usual temperature
reduction factor\cite{david}, $R_{Y}(z)=J_{0}(z)$ the orientation (Yamaji)
factor noticed in \cite{nakano99}, $R_{D}=e^{-2\pi r\Gamma _{b}/\omega _{b}}$
the Dingle exponential damping due to collision broadening of the Landau
minibands, $R_{s}={\cos \pi r\gamma }_{b}$ the spin reduction factor, which
all strongly depend on the orientation of the field, angle $\theta $ (Fig.
1). As usual, one can view the Dingle factor as corresponding to the
effective temperature $T+T_{D}$ in a clean system$,$ where the Dingle
temperature $T_{D}=\Gamma_b /\pi $. For the magic angles, where $D_{\alpha }=0,$
one has $R_{Y}=1$ and hence a pure 2D\ situation is recovered\cite{ABprb01}.
However, generally in a multiband case one should expect that only a
spectrum of a particular band will become 2D at some particular magnetic
field tilt angle.
%
%
\begin{figure}[t]
\epsfxsize=3.5in 
\epsffile{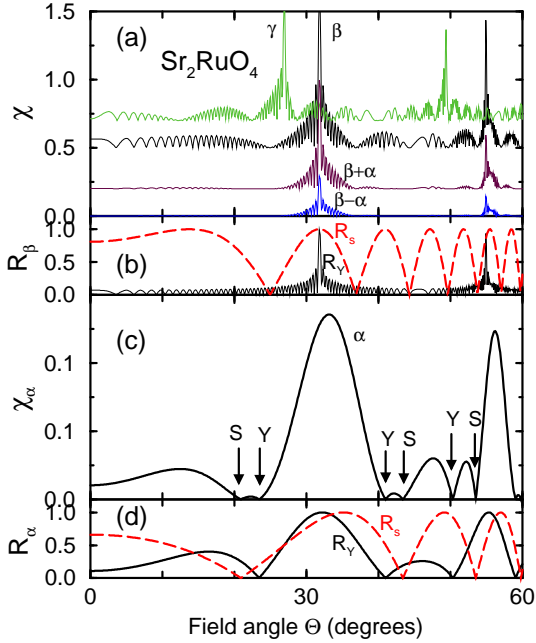 }
\vspace{.2in}
\caption{
The amplitudes of magnetic susceptibility harmonics $\protect\chi _{\protect\alpha
}$ [panel (c)]$,$ $\protect\chi _{\protect\beta },$ $\protect\chi _{\protect%
\gamma },$ and $\protect\chi _{\protect\beta \pm \protect\alpha }$
[panel (a)] in Sr$%
_{2}$RuO$_{4}$ for $T=T_{D}=0$. Parameters used in Eq.(5) are $d=6.37$\AA , $%
j=2$ for $\protect\alpha ,$ $j=1$ for $\protect\beta $ and $\protect\gamma $
bands [10]. The corresponding reduction factors $R_{Y}$ and $R_{s}$ are
shown in panels (b) for $\protect\beta $ and (d) for $\protect\alpha $
bands. The arrows mark the dips in the harmonic $\protect\chi _{\protect\alpha
}, $ corresponding to zeros of the Yamaji $R_Y$ (Y) and the spin $R_s$
(S) factors. The 
maximal amplitude of $\protect\alpha ,$ $\protect\beta ,$ and $\protect%
\beta \pm \protect\alpha $ bands is found at the first Yamaji angle 33$^{\circ }$
(see text). 
}
\label{fig:fig1}
\end{figure}

The expression for $\tilde{\Omega}$, Eq.(\ref{eq:Ot}), contains the (first)
term, responsible for the Landau diamagnetism and the oscillating term,
which is responsible for the de Haas-van Alphen effect. It is small compared
with the ``classical'' part, since $\tilde{\Omega}/\Omega _{0}\sim (\omega
/\mu )^{2}\ll 1$ for the present semiclassical situation. The Fourier
components appear with the frequencies $rF_{\alpha }$. We recover from $%
\Omega _{0}$ (\ref{eq:Om0}) and the first term in $\tilde{\Omega}$ (\ref
{eq:Ot}) the Pauli $\chi _{P}=\frac{1}{4}\mu _{B}^{2}\sum_{\alpha }\rho
_{\alpha }g_{\alpha }^{2}$ and the Landau $\chi _{L}=-\frac{1}{3}\mu
_{B}^{2}\cos ^{2}\theta \sum_{\alpha }\rho _{\alpha }(m/m_{\alpha })^{2}$
susceptibilities, respectively. Since usually $g_{\alpha }=2,$ those are
almost standard, apart from the $\cos ^{2}\theta $ factor characteristic of
the present geometry with the tilted magnetic field.

Note that the {\em chemical potential} (and carrier density) {\em oscillates}
in a closed system and, unlike in {\rm GCE,} the ``classical'' part of $%
\Omega $ contributes to oscillations as well\cite{ABprb01}. The relevant
thermodynamic potential of the closed system (CE) is the free energy, $%
F=\Omega +\mu N,$ for a fixed number of electrons, $N=-\partial \Omega
/\partial \mu $. The chemical potential is \cite{ABprb01} 
\begin{eqnarray}
\mu &=&\mu _{0}+\tilde{\mu},\\
\tilde{\mu}&=&{-}\frac{\tilde{N}}{\rho }{%
\equiv -}\frac{\tilde{N}}{\rho _{q}+\rho _{bg}}
\end{eqnarray}
where $\mu _{0}$ is the non-oscillating, while $\tilde{\mu}$ is the
oscillating part of the chemical potential, $\tilde{N}=-\partial \tilde{%
\Omega}/\partial \mu $ is the oscillating part of the density of electrons, $%
\rho =\sum_{\alpha }\rho _{\alpha }\equiv \rho _{q}+\rho _{bg}$ is the total
density of states, including the quantized $\rho _{q}$ and the background\ $%
\rho _{bg}$ DOS. Substituting this expression into $\Omega _{0}$, Eq.~(\ref
{eq:Om0}), we obtain $F=F_{0}+\tilde{F},$ where the oscillating part is \cite
{ABprb01} 
\begin{equation}
\tilde{F}=\tilde{\Omega}-{\tilde{N}^2 \over 2\rho },
%
\label{eq:Ft}
\end{equation}
while $F_{0}$ is the non-oscillating part. One sees that the difference
between the free energies of the ensembles is directly proportional to the
fluctuation of the particle density, as it should, and is suppressed by the
background density of states $\rho _{bg}$, 
$\sim \tilde{N}^2/(\rho_q+\rho_{bg})$. The oscillating part of the 
particle density is small, but both terms in (\ref{eq:Ft}) give comparable
contribution to the magnetization oscillations. Indeed, $\tilde{N}%
=\sum_{\alpha }\tilde{N}_{\alpha },$ and at low temperatures 
\begin{equation}
\frac{\tilde{N}_{\alpha }}{N_{0}}\sim \frac{\rho _{\alpha }}{\rho }\frac{B}{%
F_{\alpha }}\ll 1,\qquad T<\omega _{\alpha },
\end{equation}
since $F_{\alpha }\gg B,$ and it falls off exponentially with temperature, $%
\tilde{N}_{\alpha }/N_{0}\sim (T/\mu )\exp \left( -2\pi ^{2}T/\omega
_{\alpha }\right) ,$ at $T>\omega _{\alpha }.$ Since the density
oscillations are small, sometimes they can be screened by the back electrode
in e.g. quantum well structures \cite{shepherd99}. In a more explicit form
one obtains 
\begin{eqnarray}
\tilde{F} &=&{\frac{1}{{24}}}\sum_{\alpha }\rho _{\alpha }\omega _{\alpha
}^{2}+4\sum_{b}\sum_{r=1}^{\infty }A_{b}^{r}\cos 2\pi r\left( {\frac{F_{b}}{{%
B}}-}\frac{1}{2}\right)  \nonumber \\
&&-16\sum_{b,b^{\prime }}\sum_{r,r^{\prime }=1}^{\infty }C_{bb^{\prime
}}^{rr^{\prime }}\sin 2\pi r\left( {\frac{F_{b}}{{B}}}-\frac{1}{2}\right)
\sin 2\pi r^{\prime }\left( {\frac{F_{b^{\prime }}}{B}}-\frac{1}{2}\right) .
\label{eq:Ftil}
\end{eqnarray}
It is the last term, which yields the combination Fourier harmonics with the
frequencies $F=rF_{b}\pm r^{\prime }F_{b^{\prime }}$. Their amplitudes, 
\begin{equation}
C_{bb^{\prime }}^{rr^{\prime }}={\frac{2\pi ^{2}rr^{\prime
}A_{b}^{r}A_{b^{\prime }}^{r^{\prime }}}{\left( {\rho }_{q}+\rho
_{bg}\right) {\omega _{b}\omega _{b^{\prime }}}}}  \label{eq:Cr}
\end{equation}
are comparable with the standard single-band harmonics at low temperatures, $%
T<\omega _{\alpha }/2\pi ^{2}r$, as found earlier \cite{ABprb01} and
confirmed experimentally \cite{shepherd99,yoshida98,ohmichi99}. The
combination harmonics are suppressed in presence of the background density
of states. The spin factor depends on the tilt angle, and this results in
a strong angular dependence of the corresponding dHvA amplitudes \cite
{yoshida98,ohmichi99}. Incidentally, if one of the angular dependent factors
vanishes for some particular harmonic, $r_{0}F_{b},$ it would {\em not} mix
up with other bands to produce combination harmonics, whereas $rF_{b}$ ($%
r\neq r_{0})$ would (cf. numerical results \cite{kishigi}).

Similar to the usual 3D situation, the oscillations produce the (partial)
contributions to magnetization, $M_{\alpha },$ and susceptibility, $\chi
_{\alpha },$ which are much larger than the non-oscillating contributions.
Indeed, the ratio of corresponding amplitudes at low temperatures is $\left|
\chi _{\alpha }^{r}\right| /\chi _{0}\sim \left( F_{\alpha }/B\right)
^{2}\left( \rho _{\alpha }m/\pi ^{2}\rho |m_{\alpha }|r\right) J_{0}\left(
2\pi rD_{\alpha }/\omega _{\alpha }\right) e^{-2\pi r\Gamma _{\alpha
}/\omega _{\alpha }}\gg 1$, since $F_{\alpha }/B\gg 1$. Proportionality to $%
\left( F_{\alpha }/B\right) ^{2}$ is the property of the two-dimensional
geometry \cite{david}. The amplitudes of the standard and the combination
harmonics can be easily found from the expressions given above. The ratio of
susceptibilities is 
\begin{equation}
\frac{\chi _{_{rF_{b},\pm r^{\prime }F_{b^{\prime }}}}^{c}}{\chi
_{rF_{b}}^{\mu }}=\frac{8\pi ^{2}rr^{\prime }A_{b^{\prime }}^{r^{\prime
}}\cos \pi r^{\prime }\gamma _{b^{\prime }}}{\rho \omega _{b}\omega
_{b^{\prime }}}\left( \frac{F}{rF_{b}}\right) ^{2}  \label{eq:chir}
\end{equation}
with $F=rF_{b}\pm r^{\prime }F_{b^{\prime }}.$ Thus, for $r=r^{\prime }=1$
we have, for a warped 2-band cylindrical Fermi surface without the background
DOS, 
\begin{eqnarray}
{\frac{\chi _{F_{b},\pm F_{b^{\prime }}}^{c}}{\chi _{F_{b}}^{\mu }}} &=&{%
\frac{{4\pi }^{2}m_{b^{\prime }}}{{m_{b}+m_{b^{\prime }}}}}\frac{T{J}%
_{0}\left( 2\pi D_{b^{\prime }}/\omega _{b^{\prime }}\right) \cos \pi \gamma
_{b^{\prime }}e^{-2\pi \Gamma _{b^{\prime }}/\omega _{b^{\prime }}}}{\omega
_{b}\sinh \left( 2\pi ^{2}T/\omega _{b^{\prime }}\right) }  \nonumber \\
&&\times \left( {\frac{F_{b}\pm F_{b^{\prime }}}{F_{b}}}\right) ^{2},
\label{eq:chi11r}
\end{eqnarray}
where $\chi ^{\mu }$ denotes the amplitude of the standard dHvA harmonic for
an open system, with $\chi ^{c}$ the amplitude of the novel combination
harmonics. We see that the combination harmonics are suppressed
exponentially with respect to temperature\cite{ABprb01} and/or disorder
compared to the standard harmonics, Fig. 2. In addition, they contain the
product of the angular dependent reduction factors and, therefore, vary rapidly
with the angle $\theta $, their frequency being dominated by the band with the
largest ratio $2t_{b}/\omega _{b}$ (e.g. $\beta -$band in $\beta \pm \alpha $
harmonics in Sr$_{2}$RuO$_{4},$ Fig. 1).
%
%
\begin{figure}[t]
\epsfxsize=3in 
\epsffile{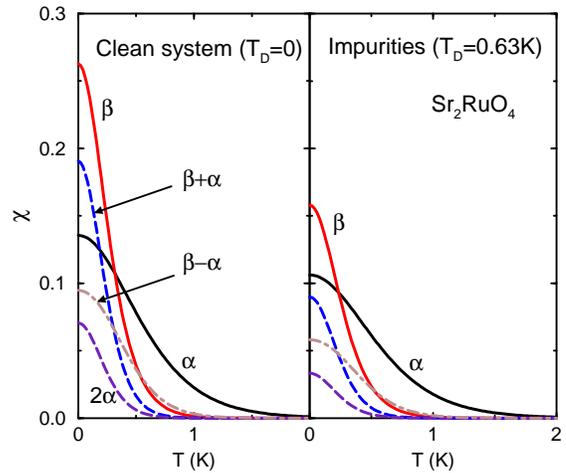}
\caption{
The amplitudes of the harmonics in Fig.~1, including also $2\alpha$,
versus the temperature in a clean (left panel) and a disordered (right 
panel) Sr$_{2}$RuO$_{4}$ at the (Yamaji) angle $\theta=33^\circ$. The
Dingle temperature $T_{D}=0.63$K [10]. 
}
\label{fig:fig2}
\end{figure}

It is important to note that even in a simplest case of a {\em one}-band
system the standard harmonics $rF_{b}$ are modified by the second quadratic
term, therefore 
\begin{eqnarray}
&&\chi _{rF_{b}}^{N}=\chi _{rF_{b}}^{\mu }+\sum_{p=1}^{r-1}\chi
_{(r-p)F_{b},pF_{b}}^{c}+\sum_{p=1}^{\infty }\chi _{(r+p)F_{b},-pF_{b}}^{c},
\label{eq:chiNd} \\
&&\chi _{rF_{b}\pm r^{\prime }F_{b^{\prime }}}^{N}=\chi _{rF_{b},\pm
r^{\prime }F_{b^{\prime }}}^{c},\quad b\neq b^{\prime },  \label{eq:chiNo}
\end{eqnarray}
and the wavefront of magnetization of the main harmonics in the closed system with
the carrier density $N={\rm const}$ is inverted compared to the open system, $\mu ={\rm %
const,}$ as it should \cite{david,alebra,alebra1}.

The present formalism is applied to Sr$_{2}$RuO$_{4}$ with the results shown
in Figs. 1 and 2, calculated with the parameters from Refs \cite
{yoshida98,andy96} and the g-factor $g_{b}=2$ for all the bands. The
main {\em maxima} in the Fourier harmonics of susceptibility $\chi $ for the
band $\alpha $ are found at $\theta _{\alpha 1(2)}=33^{\circ }$ ($55^{\circ
}),$ for the band $\beta $ at $\theta _{\beta }\approx \theta _{\alpha },$
and for band $\gamma $ at $\theta _{\gamma 1(2)}=27^{\circ }$ ($49^{\circ
}), $ which is in fair agreement with the data $\theta _{\alpha
1}=30.6^{\circ }, $ $\theta _{\beta 1}=30^{\circ }$ and $\theta _{\gamma
1}=15.3^{\circ },$ Ref. \cite{yoshida98}, and $\theta _{\alpha 1}=30^{\circ }
$, $\theta _{\beta 1(2)}=26^{\circ }~$(56$^{\circ }),$ Ref.~\cite{ohmichi99}%
. Those {\em maxima} in the harmonics apparently coincide with the points
where the Yamaji factor is unity, $R_{Y}=1,$ and the {\em system effectively
becomes 2D, }Figs.~1(b),(d). The {\em minima}, observed in $\alpha $ band at
24$^{\circ }$ \cite{yoshida98} (25$^{\circ }$ \cite{ohmichi99}$)$ and 40$%
^{\circ }$, are very close to the points were {\em both} the Yamaji $R_{Y}$
and the spin $R_{s}$ factors vanish: $R_{Y}=0$\ at 23$^{\circ }$ and 41$%
^{\circ }$, and $R_{s}=0$ at 21$^{\circ }$ and 43$^{\circ }$, Fig. 1(d).

The conspicuous rapid variation with the {\em field angle} $\theta $ of the
main, $\chi _{\beta },$ $\chi _{\gamma },$ and the combination, $\chi _{\beta
-\alpha },$ $\chi _{\beta -\alpha },$ harmonics, Fig.1(a), is explained by
the large factor $2t_{b}/\omega _{b}$ in the argument of the Bessel function
in the Yamaji factor, Eqs.(\ref{eq:Ab}), (\ref{eq:chi11r}): $2t_{\beta
}/\omega _{\beta }=7.5,$ $2t_{\gamma }/\omega _{\gamma }=5$. This
necessarily leads to the high frequency variations of the amplitudes, and
apparently only the envelope of those oscillations has been sampled
experimentally in Refs.\cite{yoshida98,ohmichi99} (see also a numerical study 
\cite{nakano99}). Those rapid variations with angle $\theta $ might be
observable, since the condition for the Yamaji approximation, $\pi
k_{fb}^{2}(t_{b}/\epsilon _{f})^{2}\ll \Delta S$ \cite{yamaji} seems to hold
for the $\beta -$band $[\pi k_{f\beta }^{2}(t_{\beta }/\epsilon
_{f})^{2}:\Delta 
S\approx 1:5],$ while for the $\gamma -$band the ratio is only about $1:3$.
Interestingly, the spin-factor $R_{s}$ defines the envelope of the $\beta -$%
amplitude, cf. Figs. 1(a) and 1(b), so the studies of the minima and maxima
on the $\chi \left( \theta \right) $ \cite{yoshida98} should allow for
accurate determination of the g-factors $g_{b}.$

The combination harmonics contain the {\em extra temperature and the Dingle
reduction factors}, so they are falling off with either temperature or
disorder, or both, faster than the standard harmonics do, see Eqs.(\ref
{eq:Cr}),(\ref{eq:Ab}), Fig.2. The temperature
dependence of the combination harmonics generally cannot be characterized by
some effective mass $m_{rb,\pm r^{\prime }b^{\prime }},$ although at higher
temperatures the relation $m_{rb,\pm r^{\prime }b^{\prime }}\approx
rm_{b}+r^{\prime }m_{b^{\prime }}$ holds approximately. All the harmonics
decay quickly with $T,$ especially in the presence of even weak disorder
(the Dingle temperature $T_{D}=0.63$K\cite{yoshida98}), Fig. 2(b), which is
in very good agreement with experiment (cf. Fig. 2 in Ref.\cite{andy96}).

Importantly, in the first experiments on Sr$_{2}$RuO$_{4}$ the disorder was
actually larger, since the elastic mean free path was $l\sim 10^{3}$\AA\ 
\cite{andy96}, compared to $l\sim $2050-5000\AA\ in the later experiments \cite
{yoshida98,ohmichi99}. The higher disorder in the first samples, and
relatively small magnitude of the combination peaks at the field $B$ directed
along the c-axis, $\theta =0,$ instead of the ``magic'' angle $\theta
=30^{\circ }$, where {\em all} the amplitudes are enhanced, Fig. 1, has
possibly prevented the discovery of the combination harmonics, predicted in Ref. 
\cite{alebra}, in the first experiments on Sr$_{2}$RuO$_{4}$\cite{andy96}.
The combination harmonics have been detected in Refs. \cite
{shepherd99,yoshida98,ohmichi99}. The dependence of the dHvA combination
amplitudes on angle, temperature, and disorder provide additional valuable
tool for studying the band structure and carrier densities in the multiband
quasi-two-dimensional metals.

\end{document}